# Enhancement in Power Conversion Efficiency of CdS Quantum Dot Sensitized Solar Cells Through a Decrease in Light Reflection

Farzaneh Ahangarani Farahani[1*]; Atila Poro[1†]; Maryam Rezaee[1]; Mehdi Sameni[2]

[1]The International Occultation Timing Association Middle East section, Iran
[2]Physics department, Payame Noor University, Arak, Iran

**Abstract**
In this research, the effect of Magnesium Fluoride ($MgF_2$) Anti-Reflection (AR) layer was investigated in quantum dot sensitized solar cells (QDSCs). $MgF_2$ nanoparticles with the dominant size of 20 nm were grown by a thermal evaporation method and a thin layer was formed on the front side of the fluorine-doped tin oxide (FTO) substrate. In order to study the effect of the AR layer on the efficiency of solar cells, this substrate was utilized in CdS QDSCs. In this conventional structure of QDSC, $TiO_2$ nanocrystals (NCs) were applied on the FTO substrate, and then it was sensitized with CdS quantum dots (QDs). According to the results, the QDSCs with $MgF_2$ AR layer represented the maximum Power Conversion Efficiency (PCE) of 3%. This efficiency was increased by about 47% compared to the reference cell without the AR layer. The reason is attributed to the presence of the AR layer and the reduction of incident light reflected from the surface of the solar cell.

## 1. Introduction

Solar energy is one of the cleanest and most accessible sources of energy, especially in areas with high annual solar radiation emissions [1-4]. So, it could be a suitable replacement for fossil fuels. Several technologies have been applied to convert solar photons into electricity. Among these solar cells are the devices to directly convert sunlight into electricity via the photoelectric effect [5-7]. QDSCs are one of the most promising types of nanostructure solar cells which have been studied here and in order recent research due to some of the most important parameters; these include good optical properties and high absorption coefficient in semiconductor QDs [8,9]. The corresponding laboratory efficiency of QDSCs is lower in reality than in theory [10]. In the past years several studies have been performed on different components in order to improve the performance of the QDSCs. This research includes applying a different structure of $TiO_2$ as the scaffold [11-13], different semiconductor QDs [14,15], electrolyte [16] and the counter electrode [17,18]. Moreover, in a solar cell the percentage of sunlight emitted is reflected which reduces the density of the generated current from photon radiation. So, another possibility to further increase the performance of QDSCs is to optimize the usage of the incident light. So far, the use of an AR layer has been used in silicon solar cells [19-23]. In order to overcome this problem, various techniques have been proposed to increase optical absorption. One of the most important techniques is to apply an AR coating [21,24,25]. The AR coatings are one of the most important components for reducing reflection from the surface; therefore, they are widely used in solar cells, lenses, and other optoelectronic devices [26-28].

As the solar cells are to be used in the air (n=1), in a case of glass where n=1.5, the material required for the AR should have an n~1.24. Such a material suitable for coating is not available, and so generally $MgF_2$ with n=1.38 is used which reduces the value of reflection from a single surface [29-31]. $MgF_2$ is a technically very important low index material in UV-VUV optical coating applications due to its high transparency and low optical loss in this wavelength range [31,32]. Therefore, $MgF_2$ can be used as an appropriate choice as an AR layer in solar cells. The most important parameters in the application of the AR layer are the number and the thickness of them rather than depending on the range of wavelengths that needs to be covered; the number of AR layers can be variable. But typically the AR coating is fabricated by using a single layer coating of a material, or a doubled layer of two selected materials [33-35].

---

[*]First author: E-mail address: ahangarani_f_1990@yahoo.com
[†]Corresponding author: E-mail address: info@iota-me.com

The current work describes MgF$_2$ composed of a single layer that has been chosen to prepare the AR layer in a standard structure of QDSCs for the first time. In the following a layer with about 8μm thickness of TiO$_2$ NCs with a dominant size of 20 nm was hydrothermally grown and deposited on FTO-glass substrates. CdS NCs were also applied for the sensitization of photoanodes utilizing the successive ionic layer adsorption and reaction (SILAR) method. The photoelectrode with a MgF$_2$ AR layer demonstrated the maximum PCE of about 3%. This efficiency was increased by about 47% compared to that of the reference cell sensitized without the AR layer.

## 2. Experimental
### 2.1. Coating of MgF$_2$ nanoparticles

In this research as the first step we needed to insure proper adhesion of the films. FTO coated glass substrates were cleaned with acetone and ethanol in an ultrasonic bath. The outer surfaces of transparent substrates were then coated with a thin layer of MgF$_2$ nanoparticles by a thermal evaporation method under a base pressure of 1×10$^{-5}$ mBar. The substrate temperature was varied from ambient temperature to a maximum value of 200°C. The evaporation rate was maintained constant at 3 Å/s with a reaction time of about 15 minutes.

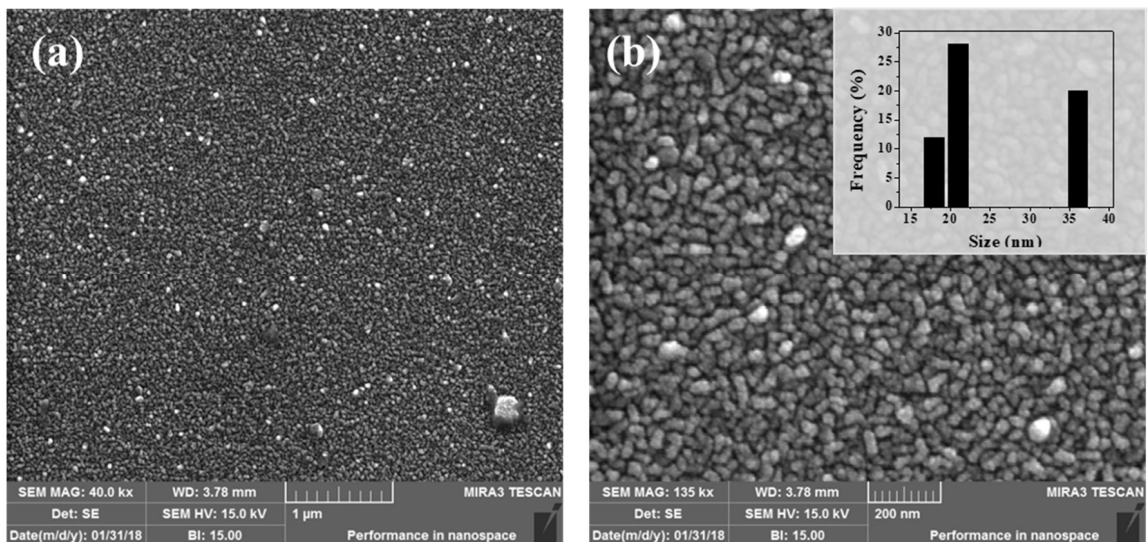

Figure 1. (a,b) Surface FESEM images of the composed layer of MgF$_2$ nanoparticles under two different magnifications. The size distribution histogram of the MgF$_2$ nanoparticles is shown in the inset of (b).

In addition to the thermal evaporation [36,37], various deposition methods have been applied for preparing MgF$_2$ thin films including Ion beam sputtering [38], e-beam evaporation [39], sol-gel processing [40], Ion-Beam Assisted Deposition (IBAD) [41], Plasma Ion Assisted Deposition (PIAD) [42].

## 3. Result and discussion
### 3.1. Coating of MgF$_2$ nanoparticles

In Figure 1(a,b) a surface Scanning Electron Microscope (SEM) view of a MgF$_2$ thin film is shown in two different magnifications. The homogeneous film consists of globular grains with diameters between 18 nm and 35 nm, as shown inset of Figure 1(b). Figure 1(a) shows that the substrate surface is completely covered with MgF$_2$ nanoparticles.

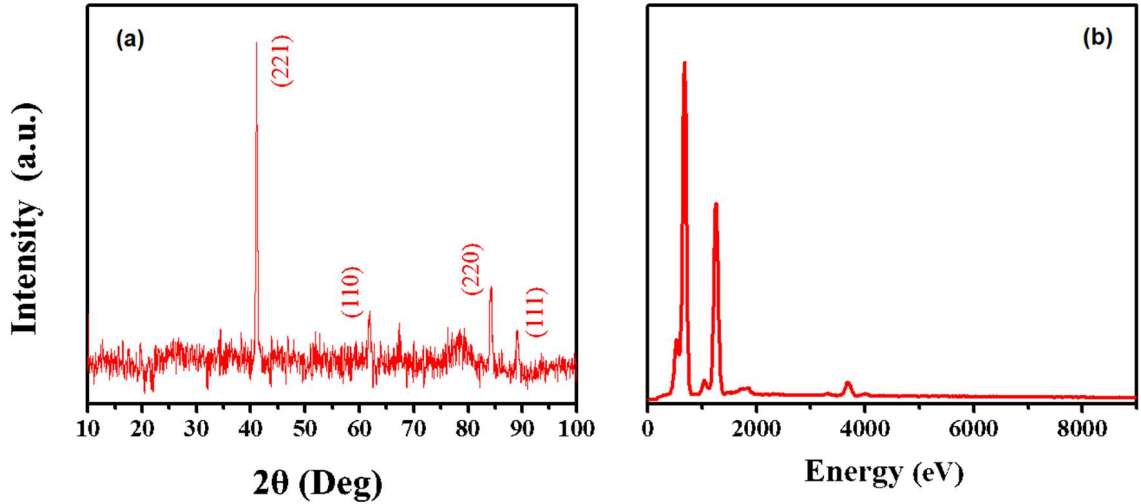

**Figure 2.** X-ray diffraction (a) and EDX (b) patterns of MgF$_2$ composed layer.

The crystalline phase of the MgF$_2$ nanoparticles layer can be detected by X-ray diffraction (XRD), as presented in Figure 2(a). The positions of XRD peaks of the MgF$_2$ films, which are observed at 41.05°, 61.83°, 84.16° and 89.16° confirm the tetragonal crystal structures of the deposited MgF$_2$ films. The obtained peaks were matched with ICSD cards No. 01-072-2231 correspond to the (221), (110), (220) and (111).

The grain size (D, nm) of the MgF$_2$ films was calculated from the main intensity peak with the following Scherer equation [43]:

$$D = \frac{(0.9 \times \lambda)}{(\beta \cos\theta)} \qquad (1)$$

where β is the Full Width at Half Maxima (FWHM), θ is the Bragg angle and λ is the wavelength of incident X-ray (Cr Kα, 0.229 nm in the present case). It is well known that the Scherer formula is a good approximation for a spherical crystal; the size is inversely proportional to the FWHM [44]. By the use of the Scherer formula for (221) plane, the average grain size of 22 nm for a MgF$_2$ film was confirmed.

Energy-Dispersive X-ray (EDX) spectroscopy analysis has been used to investigate the elemental components of the anti-reflection layer. As seen in Figure 2(b), the intensities of the peaks show the formation of both Mg and Flor atoms as the main element in the EDX pattern. The atomic composition is also calculated and shown in inset of Figure 2(b).

To analyze the quality of the produced MgF$_2$ composed layer, the optical properties have been considered. Figure 3(a) shows the measured transmittance spectra of the single-layer MgF$_2$ films deposited on a glass substrate that are compared with the spectra of uncoated glass samples. As a result, there is no noticeable change in the transmittance rate of light from either the uncoated glass or MgF$_2$ coated glass in the wavelength range of 300 nm to 1000 nm; this indicates that the MgF$_2$ composed layer is as transparent as the glass.

The diffuse reflectance spectra of the MgF$_2$ layer is also shown in Figure 3(b). As we can see, the results are matched with the transmission spectra. According to the spectra, the level of reflection from the uncoated glass surface is about 4%, while the level for the MgF$_2$ layer has been reduced to less than 1%. This result shows the suitable optical performance of the MgF$_2$ AR layer.

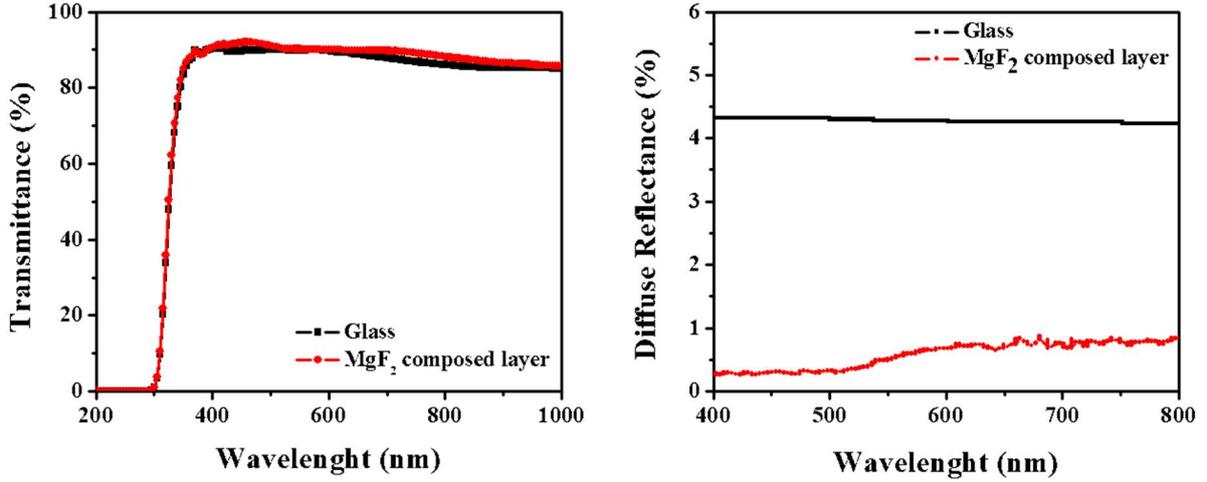

Figure 3. Transmittance (a) and diffuse reflectance (b) spectra of MgF$_2$ composed layer and uncoated layer.

In many optical devices there are several levels of separation demonstrating that the loss of intensity resulting from reflections can be serious [45]. For example, for the near-vertical radiation, the surface reflection from the crystal glass (in the air) is:

$$R_s = R_p = \left[\frac{n-1}{n+1}\right]^2 = \left[\frac{1.5-1}{1.5+1}\right]^2 = 0.04 \quad (2)$$

So for incident light, there is about 4% reflection, which will reach 6% for a flint glass with a refractive index of 1.67.

But in this work for a glass with a refractive index of 1.5 and a specific thickness of the MgF$_2$ AR layer, the magnitude of the reflection will be significantly reduced by the following equation:

$$\Delta = 2n_f t \cos\theta' \rightarrow 2n_f t = \tfrac{1}{2}\lambda \rightarrow t = \frac{\lambda}{4n_f} \quad (3)$$

$$, n_f = \sqrt{n_a n_g} \qquad R_s = \left[\frac{n_f - n_a}{n_f + n_a} - \frac{n_g - n_f}{n_g + n_f}\right]^2 \quad (4)$$

Since $n_f = 1.38$ and $n_f = 1.5$, the reflectivity is reduced from 4% to 1.3%.

All three transmission, reflection and absorption coefficients have to be considered. In the above destructive interference, according to the expected calculations, the reflection has significantly decreased.

According to the conservation of energy and the following formula:

$$A + T + R = 1 \quad (5)$$

By measuring the transmission and the small changes observed in Figure 3(a), a part of the radiation is absorbed and resulted in increasing the efficiency of the solar cell.

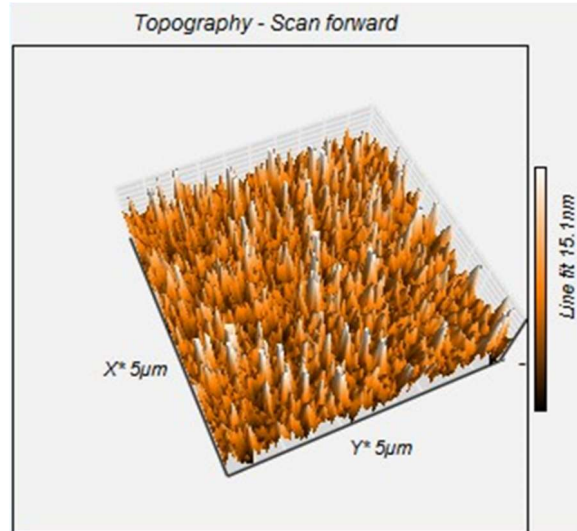

**Figure 4. The 3-Dimensional AFM image of the MgF$_2$ AR layer.**

An atomic force microscopy (AFM) image of the MgF$_2$ composed layer has been taken to measure surface topography of the layer (Figure 4). The typical root-mean-square roughness ($R_q$) and the average roughness of the layer were found at 87.993 nm and 87.91 nm, respectively. According to the average roughness, it could be observed that the MgF$_2$ surface layer was uniformly applied. This low-level roughness reduces the scattering of incident light to the surface. It does not create an imbalance in the performance of the solar cell which may also improve it.

**2.2. Fabrication of CdS quantum dot sensitized solar sells**
In this research, the photoanode includes TiO$_2$ nanoparticles as a scaffold and CdS QDs as sensitizers. So, TiO$_2$ NCs were synthesized by a hydrothermal method in acidic autoclaving PH with a Titanium Tetraisopropoxide (TTIP) precursor as described in our previous work [46]. Then the paste is prepared using TiO$_2$ nanoparticles and is deposited by the doctor blade method on the top surface of the FTO substrate. The layer was finally heated at 325, 375, 450, and 500°C for about 5, 5, 15, and 15 min to achieve a pure nanocrystalline TiO$_2$ layer. This layer is identified as the H1(2) layer.

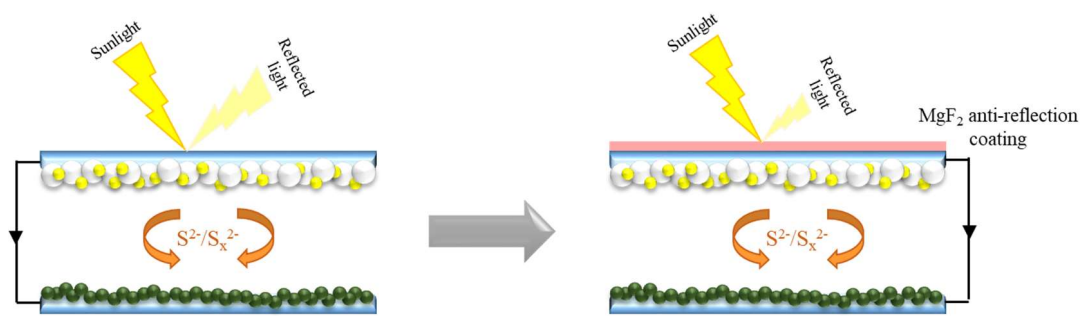

**Figure 5. Schematic of application of the anti-reflection layer in the quantum dot sensitized solar cell.**

The deposition of CdS QDs in situ on TiO$_2$ films is done by the SILAR method. For this deposition, the TiO$_2$ film was immersed in 0.1 M Cd(CH$_3$COO)$_2$.2H$_2$O methanol solution and 0.1 M Na$_2$S.9H$_2$O water/methanol solution for 1 minute, followed by rinsing with methanol and then drying. This process was repeated over six cycles as we reported earlier [47]. This layer is also identified as the H1(2)/CdS layer.
In this work, CuS NCs are used as a counter electrode that was prepared through a SILAR process by using 0.5 M Cu(NO$_3$)$_2$ in ethanol and 0.5 M Na$_2$S.9H$_2$O in ethanol/DI water solutions. The deposition was carried out to form a black layer of CuS on the surface of FTO glass substrates.

A polysulfide electrolyte solution including 1 M $Na_2S \cdot 9H_2O$, 2 M S, and 0.2 M KCl was also used. Figure 5 shows the location of the anti-reflection layer as well as its application in the quantum dot, sensitized solar cell, schematically.

## 2.3. Characterizations

All SEM images were taken using a TeScan–Mira III.EDX measurements of the photoanodes were also verified using a T-Scan field emission electron microscope. XRD patterns were measured by Philips Xpert-pro equipment with Kr $K_\alpha$ (λ= 2.29°A) radiation. Optical spectroscopies were carried out using a Mecasys Optizen POP spectrophotometer. The current-voltage characteristics were verified under AM 1.5, 100 mW/cm$^2$ simulated sun-light irradiation. The incident photon to current conversion efficiencies (IPCE) was also measured by a Sharif solar IPCE system.

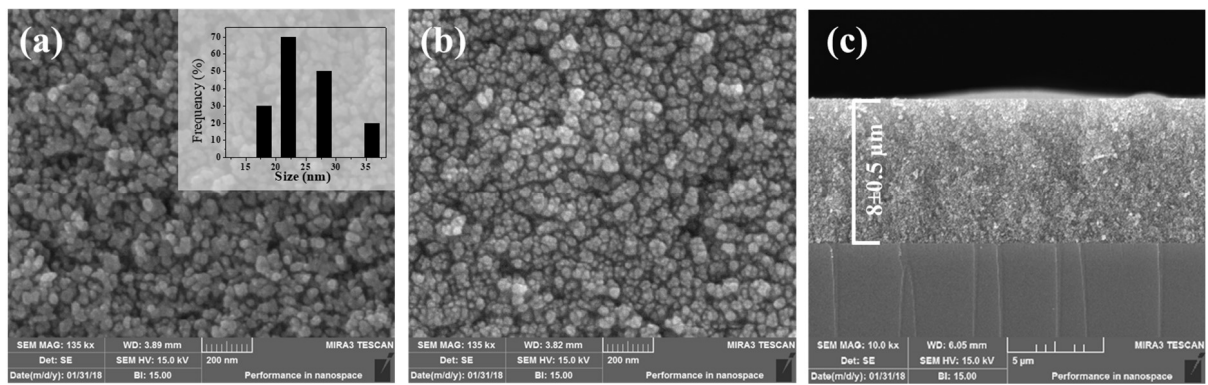

**Figure 6.** Surface FESEM pictures of $TiO_2$ NCs before (a) and after (b) deposition of CdS QDs and (c) cross-sectional SEM image of $TiO_2$ NCs layer. The inset of Figure 6(a) shows the diameter histogram of the $TiO_2$ NCs.

## 3.2. CdS sensitized solar cells

A top-view SEM image of $TiO_2$ nanoparticles-composed layer is shown in Figure 6(a). The size distribution histogram of these NCs is also shown in the inset of this Figure. It could be observed that hydrothermally grown $TiO_2$ nanoparticles with a dominate size of 20 nm are uniformly deposited on the FTO substrate. It can be seen also, that the morphology of nanocrystalline semiconductor $TiO_2$ is almost spherical in shape, and also the $TiO_2$ layer is porous that can enhance the number of QD sensitizers attached to the photoanode surface. Figure 6(b) demonstrates the top-view SEM image of the CdS sensitized $TiO_2$ NCs layer through the SILAR process after 6 cycles of deposition. According to this Figure, the surface of the $TiO_2$ composed layer is covered by a round-shape of CdS NCs. The thicknesses of the $TiO_2$ NCs layer (H1(2) layer) is also clarified in Figure 6(c) and measured about 8 μm.

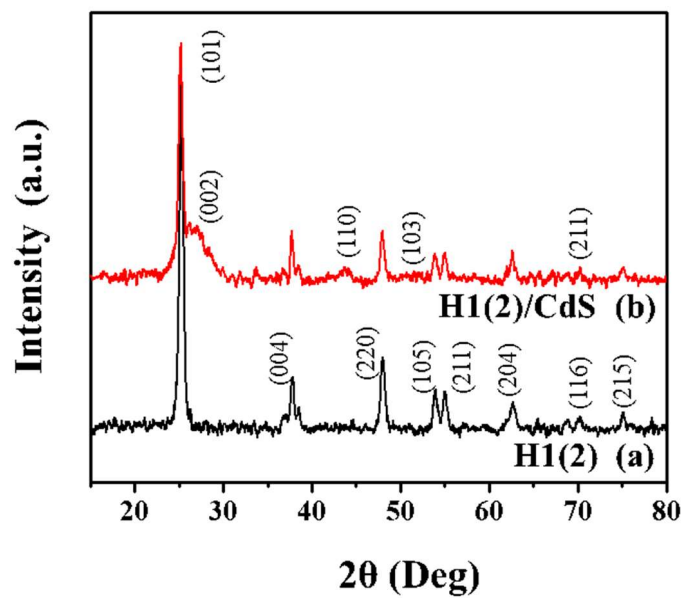

**Figure 7. X-ray diffraction patterns of the nanocrystalline H1(2) layer before and after the deposition of CdS QDs.**

Figure 7 represents the X-ray diffraction pattern of the H1(2) layer before and after CdS sensitization. As can be seen, the crystalline planes associated with the H1(2) layer in Figure 7(a) correspond to the anatase phase and belong to the standard JCPDS file No. 71.1169. The X-ray diffraction peaks are located at 2θ of 25.3°, 37.8°, 48.2°, 53.9°, 55.1°, 62.7°, 68.7°, and 75°, for $TiO_2$ nanoparticles. These peaks correspond to the (101), (004), (220), (105), (211), (204), (116), and (215) crystalline planes of the anatase phase of titanium dioxide. After deposition of CdS quantum dots these peaks were detected at 2θ of 26.65°, 43.98°, 52.25°, and 70.38°. According to JCPDS file No. 41.1049, these peaks are related to the (002), (110), (103) and (211) crystalline planes of the hexagonal phase of CdS. This plan shows that the CdS quantum dots are well deposited on the surface of the H1(2) layer.

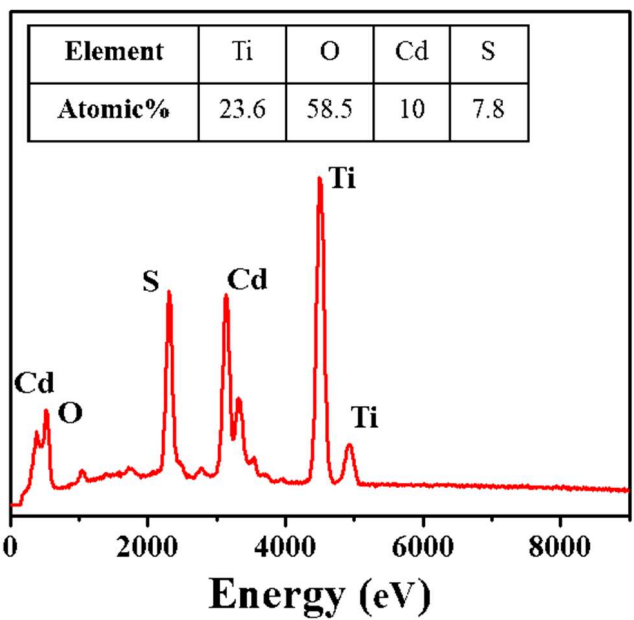

**Figure 8. EDX spectra and elemental compositions of the $TiO_2$/CdS layer.**

The composition of the CdS QDs deposited photoelectrode was determined by the EDX. According to Figure 8, Ti, O, Cd and S were detected, and their concentrations are shown by weight. These results confirmed that the CdS QDs had been deposited on the TiO$_2$ NCs layer. The EDX analysis demonstrates the purity of the sample because there is no impurity detected in Figure 8. Furthermore, the quantitate analysis shown in the inset of the Figure demonstrates the atomic composition of 23.6% Ti, 58.5% O, 10% Cd, and 7.8% S.

The optical properties of CdS QDs sensitized TiO$_2$ NCs photoelectrode is studied with UV-vis transmittance and diffuse reflectance spectroscopy as shown in Figure 9. So, Figure 9(a) presents the optical transmittance of the H1(2) layer before and after CdS sensitization over 200-1000 nm wavelength. It could be seen that the TiO$_2$ NCs layer with transparency of about 80% is highly transparent between 400 and 1100 nm. Nevertheless, after the deposition of CdS QDs, the value of transmission is considerably decreased to about 40% for the H1(2)/CdS layer. The absorption edge of the H1(2)/CdS is shifted to about 515 nm. This is due to the deposited CdS QDs on the surface of the H1(2) layer and the absorption edge is close to the bandgap energy of bulk CdS material. The diffuse reflectance spectra of fabricated photoelectrodes is also shown in Figure 9(b). This was performed in order to better determine the absorption edge of the electrode before and after the deposition of CdS QDs. Considering the diffuse reflectance spectra result, we conclude that the absorption edge starts at 500 nm for the two photoelectrodes. However, the corresponding values of diffuse reflectance are increased after the deposition of CdS QDs.

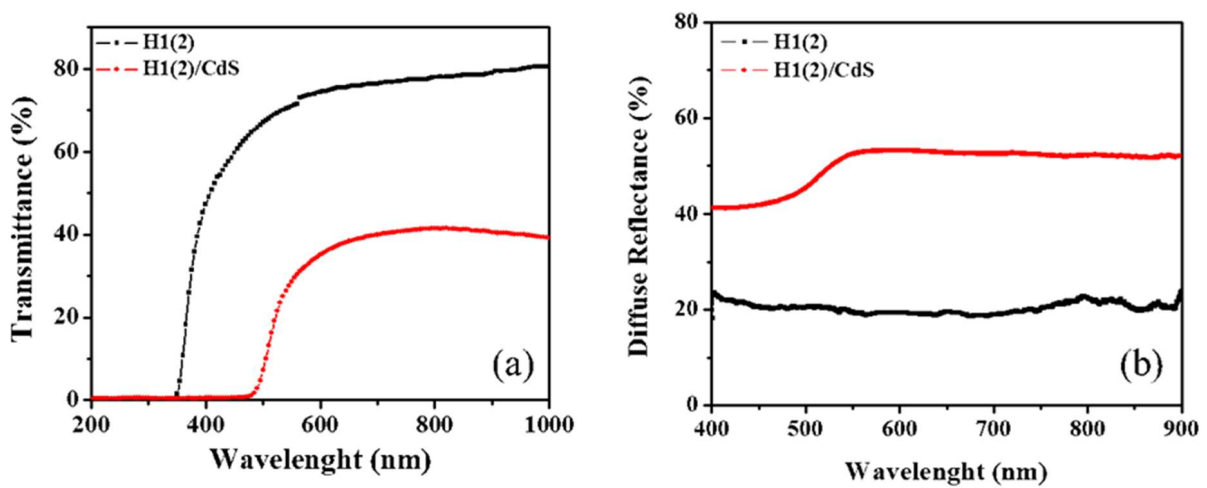

**Figure 9. Transmittance (a) and diffuse reflectance (b) spectra of H1(2) and H1(2)/CdS photoanodes.**

Finally, the photocurrent density-voltage characteristics of the cells were measured for the investigation of solar cells application and are shown in Figure 10(a). The corresponding photovoltaic parameters are also calculated and shown in Table 1. In this research, the H1(2)/CdS were selected as the reference photoanodes to show the effect of the MgF$_2$ AR layer on the performance of the CdS QDSCs. As shown in Table 1, the photovoltaic parameters of the reference cell such as short-circuit current density ($J_{sc}$), open-circuit voltage ($V_{oc}$), fill factor (FF) and conversion efficiency (η) are about 7.81 mA/cm$^2$, 558 mV, 0.47 and 2.05%, respectively. After deposition of the MgF$_2$ AR layer, these values are increased to 11.48 mA/cm$^2$, 620 mV, 0.42, and 3%. According to the result, an increase of about 47% is obvious in the short-circuit current density. It is because of the presence of the AR layer and reduction of incident light reflected from the surface of the solar cell.

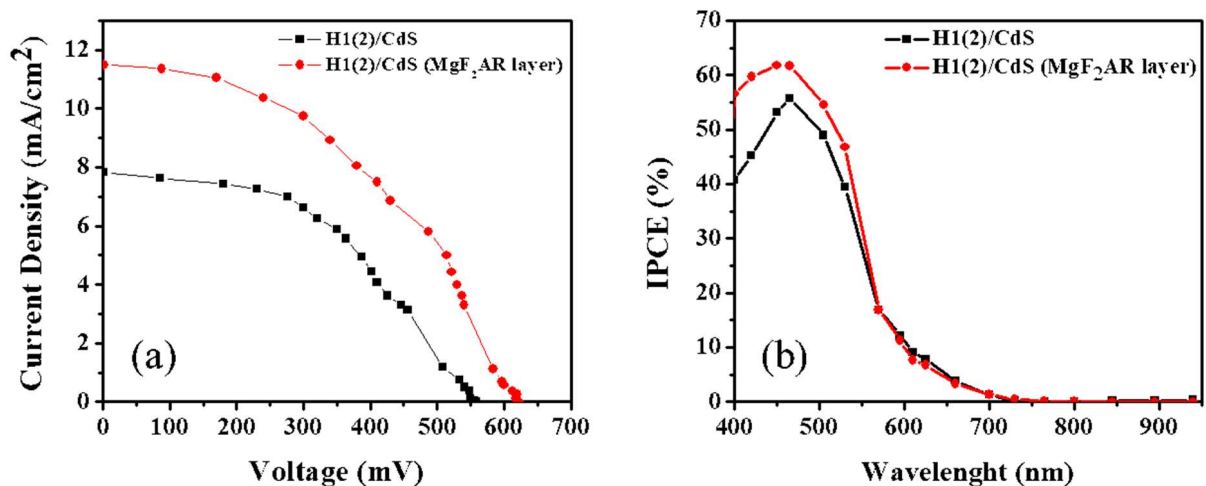

**Figure 10. The photocurrent density-voltage characteristics and the incident photon to current conversion efficiency (IPCE) spectra of the QDSCs with and without the MgF$_2$ AR layer.**

The investigation of the effect of the AR coating layer is further analyzed using IPCE measurements. Figure 10(b) shows the IPCE curves obtained for the QDSSCs fabricated with and without the MgF$_2$ AR coating layer. All the IPCE curves have a maximum of around 500 nm around the absorption edge of CdS QDs. The highest IPCE value is shown by the QDSSC fabricated with the MgF$_2$ AR layer (62%) and without the AR layer the lowest value is apparent (55.8%). This result could confirm the data of J-V measurements as were reported in Figure 10(a). The high value of IPCE supports the observation that the cell with the AR layer has excellent performance.

**Table 1. Photovoltaic parameters of the QDSCs with and without the MgF$_2$ AR coating layer.**

| Sample | Jsc | Voc | FF | η |
|---|---|---|---|---|
| H1(2)/CdS | 7.81 | 558 | 0.47 | 2.05 |
| H1(2)/CdS (MgF$_2$ AR layer) | 11.48 | 620 | 0.42 | 3 |

## 4. Conclusion

Herein, we have demonstrated the MgF$_2$ AR thin film by a thermal evaporation process. Besides, based on the SEM images and EDX analyses, the modified deposition was uniformly performed. The morphology, optical properties and the key role of the MgF$_2$ AR layer in CdS QDSCs were investigated. According to the result, the transparent MgF2 layer showed suitable optical performance that resulted in increasing the efficiency of CdS QDSCs. Moreover, the photovoltaic performances of as-prepared TiO$_2$/CdS photoanodes with an AR layer were investigated, and the results showed that this electrode exhibited higher performance than that of pure electrode. According to the results, the CdS-sensitized solar cell with a MgF$_2$ AR layer demonstrated the photovoltaic parameters of J$_{sc}$=11.48 mA/cm$^2$, V$_{oc}$=620 mV and efficiency of 3%. This efficiency was increased about 47% compared to the reference cell without the AR layer. The reason was attributed to the formation of an anti-reflection film to reduce the incident light reflection from the surface of the cell.

**Acknowledgments**

The authors would like to express their gratitude to Prof. Maziar Marandi for his advice and thanks to the Nanostructures and Nanostructured Devices research lab (NSDL) of Arak University, Iran for taking advantage of their facilities for this project. Furthermore, thanks to Paul D. Maley for providing an editorial review of the text.